\documentclass[twocolumn,prl,aps,draft,nofootinbib]{revtex4}
\usepackage{psfig}

\def\vev#1{{\langle#1\rangle}}

\def\lsim{\mathrel{\raise.3ex\hbox{$<$\kern-.75em\lower1ex\hbox{$\sim$}}}}
\def\gsim{\mathrel{\raise.3ex\hbox{$>$\kern-.75em\lower1ex\hbox{$\sim$}}}}

\begin{document}


\title{Neutrino mass limits from SDSS, 2dFGRS and WMAP}
\author{V.~Barger$^1$, Danny~Marfatia$^2$ and Adam~Tregre$^1$}
\vskip 0.3in
\affiliation{$^1$Department of Physics, University of Wisconsin, Madison, WI 53706}
\vskip 0.1in
\affiliation{$^2$Department of Physics, Boston University, Boston, MA 02215}

\begin{abstract}

We investigate whether cosmological data suggest the need for massive neutrinos. 
We employ galaxy power spectrum measurements from the Sloan Digital Sky
Survey (SDSS) and the Two Degree Field Galaxy Redshift Survey (2dFGRS), along with
cosmic microwave background (CMB) data from the Wilkinson Microwave Anisotropy 
Probe (WMAP) and 27 other CMB experiments.
We also use the measurement 
of the Hubble parameter from the Hubble Space Telescope (HST) Key Project.
We find the sum of the neutrino masses to be smaller than 0.75 eV at 2$\sigma$
(1.1 eV at 3$\sigma$).

\end{abstract}

\maketitle

Neutrino oscillation experiments provide substantial evidence that the three 
known neutrinos have 
a combined mass ($\Sigma \equiv \sum m_\nu$) 
of at least about $\sqrt{\delta m^2_{a}}\sim 0.045$ eV{\footnote{$\delta m^2_{a}$ 
is the
mass-squared difference of atmospheric neutrino oscillations. The 
mass-squared difference of solar neutrino oscillations is significantly smaller 
with $\sqrt{\delta m^2_{s}}\sim 0.008$ eV.}},
but are completely insensitive to the 
 eigenvalue of the lightest neutrino mass eigenstate~\cite{nurev}.
The neutrino mass spectrum may be hierarchical ($m_1 \ll m_2 \ll m_3$, 
$m_3 \ll m_1 \lsim m_2$) 
or quasi-degenerate ($m_1 \simeq m_2 \simeq m_3$),
depending on whether the lightest eigenmass is close to 0 or 
$\gg \sqrt{\delta m^2_{a}}$, respectively. 
The nature of the spectrum is important to neutrino mass
model-building, the contribution of
neutrinos to dark matter, and the viability of observing neutrinoless 
double beta-decay if neutrinos are Majorana~\cite{betadecay}. 

Major 
progress towards the determination of the absolute neutrino mass scale
has been made with laboratory based measurements and with
post-WMAP cosmological data. 
Tritium beta decay
experiments constrain $\Sigma$ to be smaller than
6.6 eV at 2$\sigma$~\cite{tritium} which can be improved
to $\Sigma \approx 1$~eV by the
KATRIN experiment~\cite{katrin}.
Large scale structure (LSS) data from SDSS~\cite{sdss1} combined with CMB data from 
WMAP~\cite{map} alone, yield $\Sigma \le 1.7$ eV at the 95\% C.~L.~\cite{sdss2}, 
with no strong priors or assumptions. With 2dFGRS data~\cite{2df},
CMB data and significantly stronger 
priors, a 95\% C.~L. upper limit of 1 eV was found in Ref.~\cite{hannestad}.
An even stronger bound, $\Sigma \le 0.63$ eV was obtained by the WMAP 
collaboration from a combination of 2dFGRS and CMB data and
 the 2dFGRS measurement of the galaxy bias parameter 
$b \equiv \sqrt{P_g(k)/P_m(k)}$, where $P_g$ and $P_m$ 
are the galaxy and matter
power spectra, respectively. The 2dFGRS collaboration measured a scale-independent bias 
over scales $k \sim 0.1-0.5\ h/$Mpc~\cite{bias}; 
the WMAP collaboration adopted
this bias for $k < 0.2\ h/$Mpc. A preference for massive
neutrinos found in Ref.~\cite{allen} is controversial because of the input of
a linear clustering amplitude $\sigma_8$ (defined as the rms mass 
fluctuations in spheres of radius 8 $h^{-1}$ Mpc)~\cite{sig8}; 
at present there is no experimental 
consensus on the determination of $\sigma_8$ ({\it e.g.}, see Table 5 of 
Ref.~\cite{sdss2}).  
All the above cosmological constraints were placed under the assumption
of a flat Universe in accord with the predictions of inflation.

We analyze a large set of LSS and CMB data at scales where the matter
power spectrum is linear, {\it i.e.,} $k \lsim 0.15\ h/$Mpc. 
We include the power spectrum determinations from 205,443 and 147,024 
galaxy redshifts measured by SDSS and 2dFGRS, respectively.
The CMB data comprise all WMAP data and a combination of 
the 151 band power measurements from 27 other 
CMB experiments~\cite{compilation} including
 CBI~\cite{cbi} and ACBAR~\cite{acbar}, with multipoles $l$
up to 1700 (or $k\sim 0.15\ h/$Mpc~\cite{kl}). Throughout, we
impose a top-hat prior on the Hubble constant 
$h$ ($H_0=100 h$ km/s/Mpc), from the HST~\cite{HST}. We do not include 
Ly-$\alpha$ forest data~\cite{lyman} in our analysis because 
an inversion from 
the flux power spectrum to the linear power spectrum is nonlinear and
model-dependent~\cite{seljak}.

\vskip 0.1in
\noindent
{\bf Effects of neutrino mass on the power spectrum}:

Neutrinos of eV masses are relativistic when they decouple,
and so their final number density is independent of their mass, $n_\nu=3/11
n_\gamma$. Since $\vev{E_\gamma}=2.7T_\gamma$, and $\Omega_\gamma h^2$ 
is essentially the energy density of the CMB with
$T_\gamma=2.725$ K, $n_\gamma$ is known, and
\begin{equation}
\omega_\nu\equiv \Omega_\nu h^2=n_\nu \Sigma\simeq {\Sigma \over 94.1\, {\rm eV}}\,.
\end{equation}

Neutrinos freestream on scales smaller than their Jeans length scale, which
is known as the freestreaming scale. While neutrinos
freestream, their density perturbations are damped, and simultaneously
the perturbations of cold dark matter and baryons grow more slowly because
of the missing gravitational contribution from neutrinos.
The freestreaming scale of relativistic neutrinos grows with the horizon. 
When the neutrinos become nonrelativistic, their freestreaming
 scale shrinks, they fall back into the potential wells,
and the neutrino density perturbation resumes to trace those of 
the other species.
Freestreaming suppresses
the power spectrum on scales smaller than the horizon 
when the neutrinos become nonrelativistic. 
(For eV neutrinos, this is the horizon at matter-radiation equality).
Lighter neutrinos
freestream out of larger scales and cause the power spectrum 
suppression to begin at smaller wavenumbers~\cite{morehu},
\begin{equation}
k_{nr} \simeq 0.026 \bigg({m_\nu \omega_M \over 1\, {\rm{eV}}}\bigg)^{1/2} {\rm{Mpc}}^{-1}\,,
\end{equation}
assuming almost degenerate neutrinos. Here, 
$\omega_M\equiv \Omega_M h^2$ is the total matter density (which is comprised of
baryons, cold dark matter and massive neutrinos).
On the other hand, heavier neutrinos
constitute a larger fraction of the matter budget and suppress power on 
smaller scales more strongly than lighter neutrinos~\cite{power}:
\begin{equation}
{\Delta P_m \over P_m} \approx -8 {f_\nu}\simeq 
-0.8 \bigg({\Sigma \over 1\, {\rm{eV}}}\bigg) \bigg({0.1 \over \omega_M}\bigg)\,,
\label{suppress}
\end{equation}
where $f_\nu\equiv \Omega_\nu/\Omega_M$ is 
the fractional contribution of neutrinos
to the total matter density. 

Analyses of CMB data are not sensitive to neutrino masses due to the fact that
at the epoch of last scattering, eV mass neutrinos behave essentially
like cold dark matter. (WMAP data alone allow the dark matter to be entirely constituted
by massive neutrinos~\cite{sdss2}).
However, an important role of CMB data 
is to constrain other parameters that are degenerate with $\Sigma$. Also, 
since there
is a range of scales common to the CMB and LSS experiments, CMB data provides
an important constraint on the bias parameters. Sensitivity to neutrino
masses results from the complementarity of galaxy surveys and CMB experiments.

\begin{figure}[tbh]
  \begin{center}
\psfig{file=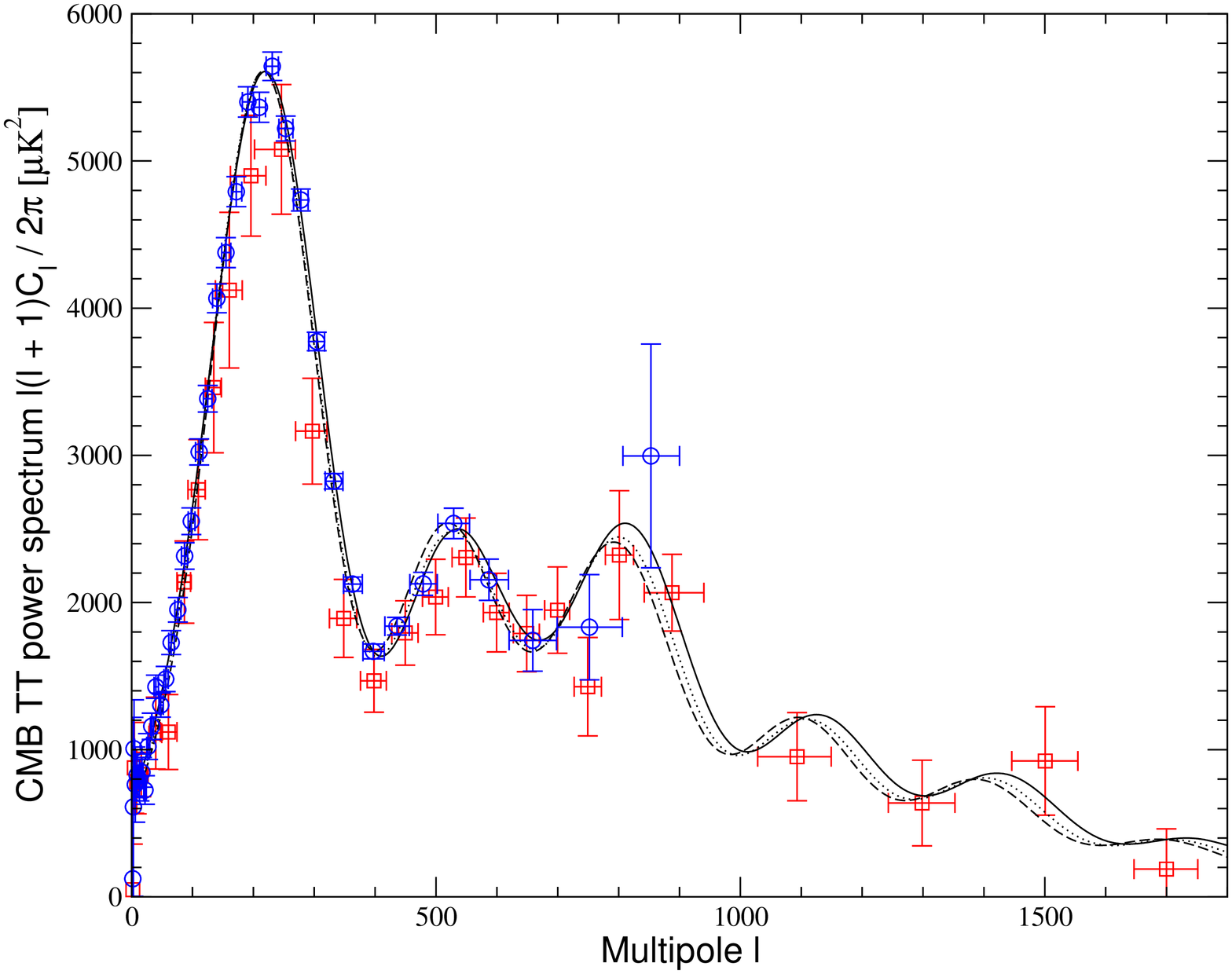,angle=270,width=8.8cm,height=6.5cm}
\psfig{file=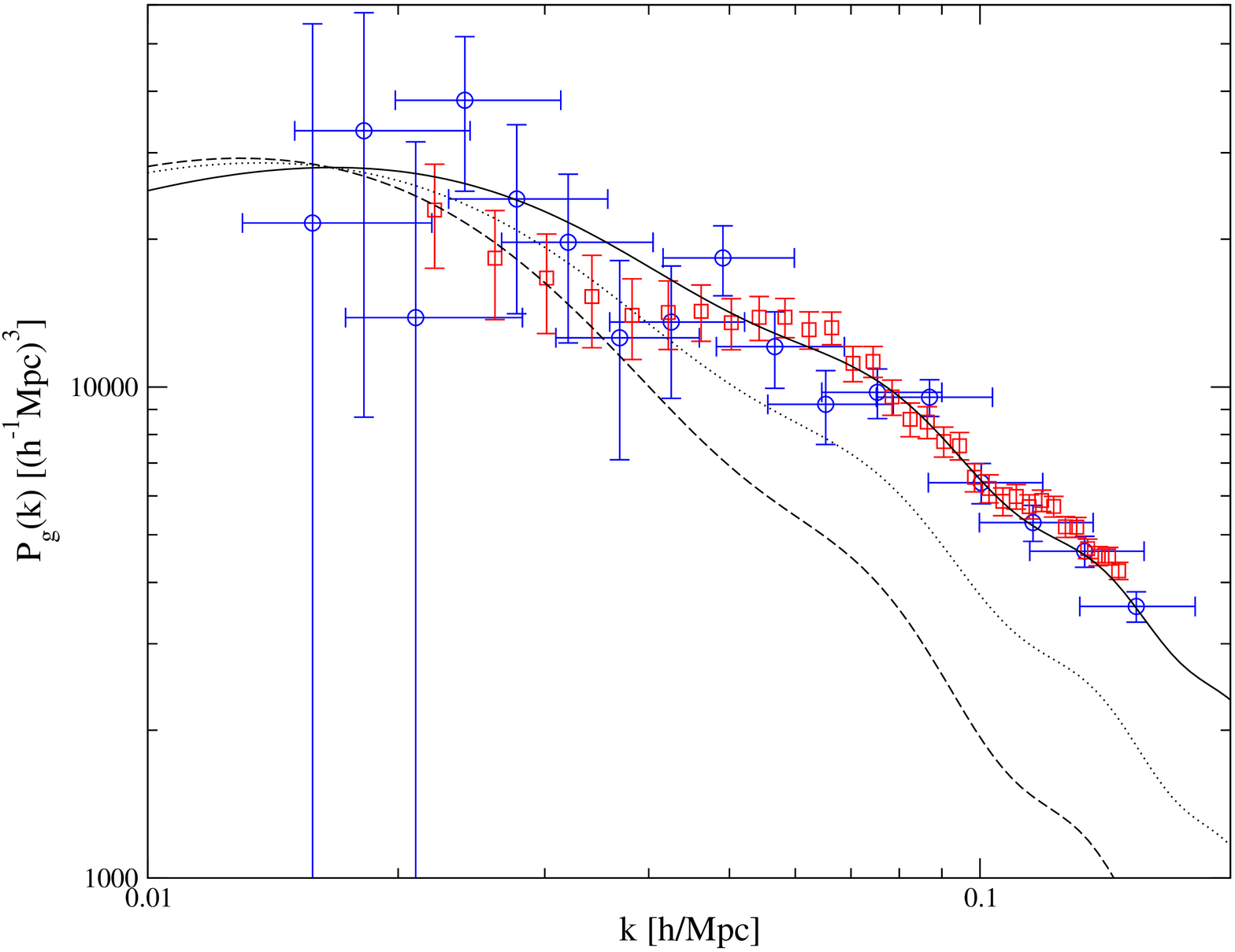,angle=270,width=8.8cm,height=6.5cm}
\caption[]{Upper panel: CMB TT power spectra. Lower panel: Galaxy power spectra. 
The curves are the spectra for $\Sigma=0.28$ eV (solid, best-fit parameters; the 
galaxy power spectrum is shown for the SDSS best-fit normalization), 
$\Sigma=1.5$ eV (dotted) and $\Sigma=3$ eV (dashed). The latter two spectra 
 have all other parameters 
(except the normalization, bias parameters and $\Omega_{cdm}$) fixed at the best-fit
values. All curves are for a flat universe.
The CMB TT spectra
are normalized to have identical powers at the first peak. The galaxy power spectra
are normalized to have identical powers at $k=0.017\ h$/Mpc. In the upper panel,
the data points marked by circles (squares) represent the binned TT spectrum 
from WMAP (pre-WMAP experiments). In the
lower panel, the data points marked by circles (squares) represent the galaxy 
power spectra from the 17 SDSS (32 2dFGRS)
bands used in our analysis.
}
    \label{fig:suppress}
  \end{center}
\end{figure}

Figure~\ref{fig:suppress} shows that the suppression of power 
caused by massive neutrinos is much greater for the galaxy power spectrum 
than for the CMB TT spectrum.  
We do not show the effect of
neutrino masses on the CMB TE spectrum because it is tiny. Note that we have
normalized the spectra to emphasize the power suppression at small
scales.

\vskip 0.1in
\noindent
{\bf Analysis}:

We compute the CMB TT and TE power spectra $\delta T_l^2=l(l+1)C_l/2 \pi$, and
the matter power spectrum $P_m(k)$, all in the linear approximation, 
using the Code for Anisotropies in the 
Microwave Background or CAMB~\cite{camb} (which is a parallelized
version of CMBFAST~\cite{cmbfast}). 
We assume 
the Universe to be flat, $\Omega_{tot}=1$, that
the dark energy is in the form of a cosmological constant $\Lambda$, and that
there are three neutrino species.
We calculate
the angular power  spectra on a grid defined by
$\omega_M$, $f_\nu$, the baryon
density $\omega_B\equiv \Omega_B h^2$, the Hubble constant
$h$, the reionization optical depth $\tau$, and the spectral index
$n_s$ of the primordial power spectrum. 

We employ the following grid:
\begin{itemize}
\item {$0.05 \leq \omega_M \leq 0.27$ in steps of size 0.02, and $\omega_M=0.14$, 0.16.}
\item {$0 \leq f_\nu \leq 0.15$ in steps of size 0.01.}
\item {$0.018 \leq \omega_B \leq 0.028$ in steps of size 0.001.} 
\item {$0.64 \leq h \leq 0.80$ in steps of size 0.02.}
\item {$0 \leq \tau \leq 0.3$ in steps of size 0.025.}
\item {$0.8 \leq n_s \leq 1.2$ in steps of size 0.02.}
\item {The normalization of the primordial power 
spectrum $A_s$, is a continuous parameter.}
\item {The bias parameters $b_{SDSS}$ and $b_{2dF}$ are 
scale-independent{\footnote{On large scales, galaxy bias is expected to be
a scale-independent constant~\cite{scaleindep}. This has been confirmed by the
2dFGRS collaboration~\cite{bias}.}} and continuous.}
\end{itemize}

The suppression of small scale power 
depends directly on $f_\nu$ and indirectly on $\Sigma$.
From Eq.~(\ref{suppress}), $\omega_M$ is strongly degenerate 
with $\Sigma$,
requiring independent knowledge of $\omega_M$ to break the degeneracy. The SDSS
collaboration only used WMAP data to provide this information and
found the somewhat weak, but conservative, 95\% C.~L. 
bound $\Sigma\le 1.7$ eV~\cite{sdss2}. 
Their analysis also gave a 1$\sigma$ constraint 
$h=0.645^{+0.048}_{-0.040}$, which lies at the lower end of the
HST measurement  $h=0.72\pm 0.08$~\cite{HST}. 
It is known that less stringent constraints on $\Sigma$ are obtained for
lower values of $h$~\cite{degen} because CMB data then allow larger 
$\Omega_M$~\cite{break}{\footnote{As a proof of principle, 
it was shown in Ref.~\cite{degen} 
that a flat non-$\Lambda$CDM model with $h=0.45$, $\Omega_M=1$
 and $\Sigma=3.8$ eV, 
provides as good a representation of the 2dFGRS and pre-WMAP 
CMB data as a flat $\Lambda$CDM model with  $h=0.7$,
$\Omega_M=0.3$ and massless neutrinos. This is true even including the WMAP data
provided different power-laws are used to describe the spectrum
for $l$ above and below the first peak~\cite{sarkar}.}}. 
Aside from the fact that we are using a significantly larger dataset than the SDSS 
collaboration, we expect our analysis to yield a stronger constraint 
on $\Sigma$ simply because we constrain $h$ 
by the HST measurement. 
Note that our grid-range for $\omega_M$ is almost identical to a 
combination of the 3$\sigma$ range of $\Omega_M$ allowed by 
SN Ia redshift data~\cite{sn}, and the HST prior on $h$. 

In our analysis, we conservatively include only the first 17 
SDSS band powers, for which $0.016 \le k \le 0.154\ h/$Mpc, 
and the power spectrum is in the linear regime.
We use the window functions and 
likelihood code provided by the SDSS collaboration~\cite{sdss1},
and leave the bias parameter $b_{SDSS}$ free.

For 2dFGRS, only 32 band powers with $0.022 \le k \le 0.147\ h/$Mpc are 
included. The window functions and covariance matrix have been made
publicly available by the 2dFGRS collaboration~\cite{2df}. 
The bias parameter $b_{2dF}$ is left free. 

The WMAP data are in the form of 899 measurements of the TT power spectrum from
$l=2$ to $l=900$~\cite{maptt} and 449 data points of the TE power
spectrum~\cite{mapte}. 
We compute the  likelihood of each model of our grid using Version 1.1 of the
code provided by the collaboration~\cite{mapcode}. The WMAP code computes 
the full covariance matrix under the assumption that the off-diagonal terms are
subdominant. This approximation breaks down for unrealistically small
amplitudes.  
When the height of the first peak is below 5000 $\mu K^2$
(which is many standard deviations away from the data), only
the diagonal terms of the covariance matrix are used to compute the likelihood.

We include the combined CMB data from pre-WMAP experiments,
by using the 28 pre-WMAP band powers, the  window functions and the correlation
matrix compiled in Ref.~\cite{compilation}.

We obtain the $x$-$\sigma$ range of $\Sigma$, by selecting those parameter 
sets with $\Delta \chi^2=\chi^2-\chi^2_{min}\le x^2$, 
after minimizing over all other parameters. 

\vskip 0.1in
\noindent
{\bf Results and conclusions}:

\begin{figure}[tbh]
  \begin{center}
\psfig{file=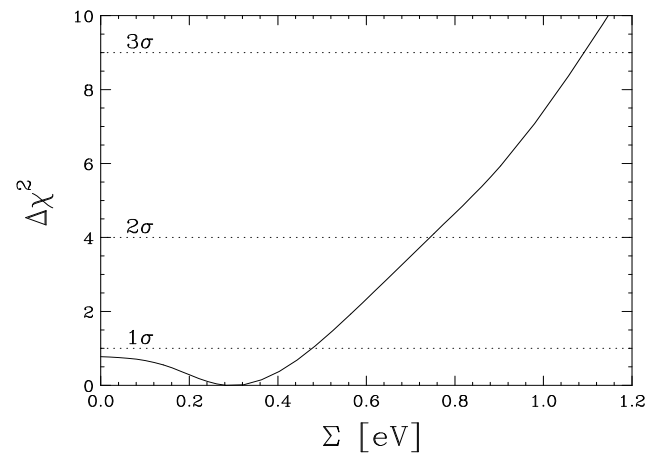,width=8.5cm,height=6cm}
\caption[]{$\Delta \chi^2$ vs $\Sigma$ from an analysis of SDSS, 2dFRGS, WMAP and 
other CMB data with $b_{SDSS}$ and $b_{2dF}$ free.
}
    \label{fig:cosmobds}
  \end{center}
\end{figure}


Our analysis yields the best-fit parameters $f_\nu=0.02$, 
$\omega_M=0.15$, $\omega_B=0.023$, $h=0.66$, $\tau=0.075$, $n_s=0.96$, 
$A_s=21.38\times 10^{-10}$
with $\chi^2=1499.83$ for $1425-9=1416$
degrees of freedom. From the normalizations of the power spectra required to fit
the SDSS and 2dFGRS data, we obtain 
$b_{SDSS}=1.13^{+0.08}_{-0.13}$ and 
$b_{2dF}=1.20^{+0.10}_{-0.13}$ (1$\sigma$ ranges).

Figure~\ref{fig:cosmobds} shows $\Delta \chi^2$ versus 
$\Sigma$. 
The neutrino mass bound is $\Sigma \le 0.75$ eV at $2\sigma$ 
($\Sigma \le 1.1$ eV at $3\sigma$). 
Thus, cosmological data do not require a 
significant neutrino dark matter component, and
are increasingly rejecting a quasi-degenerate neutrino mass spectrum.  




Eventually, lensing measurements of galaxies and the CMB by large scale structure
are expected to probe a hierarchical neutrino mass spectrum with 
$\Sigma \approx 0.04$ eV~\cite{kaplinghat}.

\vskip 0.1in
{\it Acknowledgments}:
We thank M. Tegmark for communications regarding the SDSS data, and S. Bridle and 
L. Verde for communications regarding the 2dFGRS data.
The computations were carried out on the CONDOR system at the University
of Wisconsin, Madison. 
This research was supported by the U.S.~DOE  
under Grants No.~DE-FG02-95ER40896 and No.~DE-FG02-91ER40676,
and by the Wisconsin Alumni Research Foundation. 


\begin{thebibliography}{99}


\bibitem{nurev}
For a recent review see, V.~Barger, D.~Marfatia and K.~Whisnant
Int.\ J.\ Mod.\ Phys.\ E {\bf 12}, 569 (2003)
[arXiv:hep-ph/0308123].

\bibitem{betadecay}
V.~Barger, S.~L.~Glashow, D.~Marfatia and K.~Whisnant,
Phys.\ Lett.\ B {\bf 532}, 15 (2002)
[arXiv:hep-ph/0201262].

\bibitem{tritium}
C.~Weinheimer,
arXiv:hep-ex/0210050;
V.~M.~Lobashev {\it et al.},
Nucl.\ Phys.\ Proc.\ Suppl.\  {\bf 91}, 280 (2001).

\bibitem{katrin}
A.~Osipowicz {\it et al.}  [KATRIN Collaboration],
arXiv:hep-ex/0109033.

\bibitem{sdss1}
M.~Tegmark {\it et al.}  [SDSS Collaboration],
arXiv:astro-ph/0310725.

\bibitem{map}
C.~L.~Bennett {\it et al.},
Astrophys.\ J.\ Suppl.\  {\bf 148}, 1 (2003)
[arXiv:astro-ph/0302207].

\bibitem{sdss2}
M.~Tegmark {\it et al.}  [SDSS Collaboration],
arXiv:astro-ph/0310723.

\bibitem{2df}
W.~J.~Percival {\it et al.},
arXiv:astro-ph/0105252;
M.~Colless {\it et al.},
arXiv:astro-ph/0106498.

\bibitem{hannestad}
S.~Hannestad,
JCAP {\bf 0305}, 004 (2003)
[arXiv:astro-ph/0303076].

\bibitem{bias}
L.~Verde {\it et al.},
Mon.\ Not.\ Roy.\ Astron.\ Soc.\  {\bf 335}, 432 (2002)
[arXiv:astro-ph/0112161].

\bibitem{allen}
S.~W.~Allen, R.~W.~Schmidt and S.~L.~Bridle,
arXiv:astro-ph/0306386.

\bibitem{sig8}
S.~W.~Allen, A.~C.~Fabian, R.~W.~Schmidt and H.~Ebeling,
Mon.\ Not.\ Roy.\ Astron.\ Soc.\  {\bf 342}, 287 (2003)
[arXiv:astro-ph/0208394].

\bibitem{compilation}
X.~Wang, M.~Tegmark, B.~Jain and M.~Zaldarriaga,
arXiv:astro-ph/0212417.

\bibitem{cbi}
T.~J.~Pearson {\it et al.},
arXiv:astro-ph/0205388.

\bibitem{acbar}
C.~l.~Kuo {\it et al.}  [ACBAR collaboration],
arXiv:astro-ph/0212289.

\bibitem{kl}
M.~Tegmark and M.~Zaldarriaga,
Phys.\ Rev.\ D {\bf 66}, 103508 (2002)
[arXiv:astro-ph/0207047];
D.~Scott, J.~Silk and M.~J.~White,
Science {\bf 268}, 829 (1995)
[arXiv:astro-ph/9505015].

\bibitem{HST}
W.~L.~Freedman {\it et al.},
Astrophys.\ J.\  {\bf 553}, 47 (2001)
[arXiv:astro-ph/0012376].

\bibitem{lyman}
R.~A.~Croft {\it et al.},
Astrophys.\ J.\  {\bf 581}, 20 (2002)
[arXiv:astro-ph/0012324];
P.~McDonald, J.~Miralda-Escude, M.~Rauch, W.~L.~W.~Sargent, T.~A.~Barlow, R.~Cen and J.~P.~Ostriker,
arXiv:astro-ph/9911196.

\bibitem{seljak}
U.~Seljak, P.~McDonald and A.~Makarov,
Mon.\ Not.\ Roy.\ Astron.\ Soc.\  {\bf 342}, L79 (2003)
[arXiv:astro-ph/0302571].

\bibitem{morehu}
W.~Hu and D.~J.~Eisenstein,
Astrophys.\ J.\  {\bf 498}, 497 (1998)
[arXiv:astro-ph/9710216];
D.~J.~Eisenstein and W.~Hu,
Astrophys.\ J.\  {\bf 511}, 5 (1997)
[arXiv:astro-ph/9710252].

\bibitem{power}
W.~Hu, D.~J.~Eisenstein and M.~Tegmark,
Phys.\ Rev.\ Lett.\  {\bf 80}, 5255 (1998)
[arXiv:astro-ph/9712057]. 

\bibitem{camb}
A.~Lewis, A.~Challinor and A.~Lasenby,
Astrophys.\ J.\  {\bf 538}, 473 (2000)
[arXiv:astro-ph/9911177];
http://camb.info/

\bibitem{cmbfast}
U.~Seljak and M.~Zaldarriaga,
Astrophys.\ J.\  {\bf 469}, 437 (1996)
[arXiv:astro-ph/9603033].

\bibitem{scaleindep}
A.~Taruya, H.~Magara, Yi.~P.~Jing and Y.~Suto,
PASJ {\bf 53}, 155 (2001).

\bibitem{degen}
O.~Elgaroy and O.~Lahav,
JCAP {\bf 0304}, 004 (2003)
[arXiv:astro-ph/0303089].

\bibitem{break}
J.~A.~Rubino-Martin {\it et al.},
Mon.\ Not.\ Roy.\ Astron.\ Soc.\  {\bf 341}, 1084 (2003)
[arXiv:astro-ph/0205367].

\bibitem{sarkar}
A.~Blanchard, M.~Douspis, M.~Rowan-Robinson and S.~Sarkar,
arXiv:astro-ph/0304237.

\bibitem{sn}
J.~L.~Tonry {\it et al.},
Astrophys.\ J.\  {\bf 594}, 1 (2003)
[arXiv:astro-ph/0305008].

\bibitem{maptt}
G.~Hinshaw {\it et al.},
Astrophys.\ J.\ Suppl.\  {\bf 148}, 135 (2003)
[arXiv:astro-ph/0302217].

\bibitem{mapte}
A.~Kogut {\it et al.},
Astrophys.\ J.\ Suppl.\  {\bf 148}, 161 (2003)
[arXiv:astro-ph/0302213].

\bibitem{mapcode}
L.~Verde {\it et al.},
Astrophys.\ J.\ Suppl.\  {\bf 148}, 195 (2003)
[arXiv:astro-ph/0302218].

\bibitem{kaplinghat}
W.~Hu and M.~Tegmark,
Astrophys.\ J.\ Lett.\ {\bf 514}, 65 (1999) 
[arXiv:astro-ph/9811168];
K.~N.~Abazajian and S.~Dodelson,
Phys.\ Rev.\ Lett.\  {\bf 91}, 041301 (2003)
[arXiv:astro-ph/0212216];
M.~Kaplinghat, L.~Knox and Y.~S.~Song,
arXiv:astro-ph/0303344.

\end{thebibliography}
\end{document}